\documentclass[twocolumn,floatfix, showpacs,nofootinbib]{revtex4}

\usepackage[final]{epsfig}
\usepackage{amsmath}

\begin{document}
 
\title{Energy loss in a fluctuating hydrodynamical background}
 
\author{Thorsten Renk}
\email{thorsten.i.renk@jyu.fi}
\author{Hannu Holopainen}
\email{hannu.l.holopainen@jyu.fi}
\author{Jussi Auvinen}
\email{jussi.a.m.auvinen@jyu.fi}
\author{Kari Eskola}
\email{kari.eskola@phys.jyu.fi}
\affiliation{Department of Physics, P.O. Box 35, FI-40014 University of Jyv\"askyl\"a, Finland}
\affiliation{Helsinki Institute of Physics, P.O. Box 64, FI-00014 University of Helsinki, Finland}


\begin{abstract}
Recently it has become apparent that event-by-event fluctuations in the initial state of hydrodynamical modelling of ultrarelativistic heavy-ion collisions are crucial in order to understand the full centrality dependence of the elliptic flow coefficient $v_2$. In particular, in central collisions the density fluctuations play a major role in generating the spatial eccentricity in the initial state.
This raises the question to what degree high $P_T$ physics, in particular leading-parton energy loss, which takes place in the background of an evolving medium, is sensitive to the presence of the event-by-event density fluctuations in the background. In this work, we report results for the effects of fluctuations on the nuclear modification factor $R_{AA}$
in both central and noncentral $\sqrt{s_{NN}}=200$~GeV Au+Au collisions at RHIC. Two different types of energy-loss models, a radiative and an elastic, are considered. In particular, we study the dependence of the results on the 
assumed spatial size of the density fluctuations, and discuss the angular modulation of $R_{AA}$ with respect to the event plane.
\end{abstract}
 
\maketitle

\section{Introduction}
\label{sec1}

The expression "jet tomography" is often used to describe the analysis of 
hard perturbative QCD (pQCD) processes taking place inside the soft medium created in an 
ultrarelativistic heavy-ion collision. Jet tomography with the aim to study properties 
of the medium has become one of the core observables for heavy-ion physics at RHIC, and will be
even more important in the LHC heavy-ion program due to the large kinematic reach of the LHC.
In particular, the focus so far has usually been on the nuclear modification factor $R_{AA}$ between
high-energy hadron production in $A$-$A$ collisions and the scaled expectation from p-p collisions. 
The significant suppression seen in $R_{AA}$ follows from the energy loss of the leading shower-parton caused by its interactions with the soft QCD medium, see e.g. \cite{Tomo1,Tomo2,Tomo3,Tomo4,Eskola:2004cr,Tomo5}.

Early computations of energy loss were based on rather schematic models of the medium, like static cylinders. In other words, they implicitly assumed that the mean density of the medium is the only relevant tomographical information reflected in observables. However, subsequent systematic studies of the role of the medium evolution model for energy loss \cite{SysHyd1,SysHyd2,SysHyd3} have shown that this is not the case: Both the azimuth-angle integrated $R_{AA}$ in central collisions as well as the difference between the yields of high-$p_T$ hadrons in the reaction plane and out of the reaction plane in non-central collisions reflect the details of the medium evolution dynamics \cite{SysHyd3}.

On the other hand, in recent years it has been realized that details of the bulk-medium evolution, in particular the momentum space asymmetries (driven by the different pressure gradients the in-plane and out of plane directions), which are commonly measured as the second harmonic coefficient $v_2$, can only be understood properly by taking initial state density fluctuations into account \cite{Holopainen:2010gz,Schenke:2010rr,Qiu:2011iv}. This raises a question of how the averaging over many events is best performed when the medium is described using relativistic fluid dynamics. The measured $v_2$ represents an average over many different events. Previously it was implicitly assumed that it is sufficient to average over many different initial state geometries and run a hydrodynamical simulation once given this averaged, smooth initial state. However, state of the art models now simulate the fluid-dynamical evolution event by event (EbyE) for every initial state, and only then average over the resulting particle spectra.

One may therefore also ask to what degree the order of averaging matters when considering energy loss in a hydrodynamical background medium. An exploratory, rather schematic study of this problem has been presented in \cite{Rodriguez:2010di,Fries:2010jd}. It is the aim of this paper to study the problem in the context of more detailed EbyE fluid-dynamical and energy-loss models which are constrained by a large body of data.

\section{Initial state fluctuations in hydrodynamical modeling}
\label{sec2}

We use the event-by-event ideal hydrodynamical framework presented in
Ref.~\cite{Holopainen:2010gz} to model the spacetime evolution of the bulk QCD-matter produced in
ultrarelativistic heavy ion collisions.

\subsection{Initial state}

We apply a Monte Carlo Glauber (MCG) model to produce realistic initial states
with density fluctuations. Nucleons are distributed into each nucleus
using a standard two-parameter Woods-Saxon distribution. We have not included any
finite nucleon-size effects. The colliding nuclei are separated by an impact parameter
$b$ taken randomly from a distribution $dN/db \sim b$. Nucleons $i$ and $j$ from
the different nuclei are assumed to collide if
\begin{equation}
  (x_i-x_j)^2 + (y_i-y_j)^2 \le \frac{\sigma_{NN}}{\pi},
\end{equation}
where $\sigma_{NN}$ is the inelastic nucleon-nucleon cross section and $x_i,y_i$ are
the transverse coordinates for the nucleon $i$.

The initial energy density profile $\epsilon(x,y)$ is obtained by distributing the density around
the transverse positions of the wounded nucleons (WN) or binary collisions (BC), and using
a 2-dimensional Gaussian smearing
\begin{equation}
 \epsilon (x,y) = \frac{K}{2\pi \sigma^2} \sum_{i=1}^{N_{\rm WN/BC}} \exp \Big( -\frac{(x-x_i)^2+(y-y_i)^2}{2\sigma^2} \Big),
\end{equation}
where $\sigma$ controls the width of the Gaussian. The overall normalization constant $K$, as well as
the initial time $\tau_0 = 0.17$~fm for the hydrodynamical evolution, are motivated by
the EKRT model \cite{Eskola:1999fc} as in Ref.~\cite{Holopainen:2010gz}. For the WN (BC) profile
$K = 37.8\, (12.4)$ GeV/fm. We vary the width parameter $\sigma$ in the range 0.4--0.8 fm.

Centrality classes are defined using the number of WN. We slice the distribution of the events into
intervals of $N_{\rm WN}$ so that each interval has a certain percentage of the total events.
The impact parameter varies freely in each centrality class. For simplicity we use
the same centrality selection for the WN and BC profiles.

\subsection{Hydrodynamics}
We solve ideal hydrodynamical equations $\partial_\mu T^{\mu\nu} = 0$, where
$T^{\mu\nu} = (\epsilon+P)u^\mu u^\nu - g^{\mu\nu} P$ is the stress-energy
tensor, $u^\mu$ is the fluid four-velocity and $P$ is the pressure. We simplify
the system by assuming longitudinal boost-invariance and by neglecting
the net-baryon number density. Both simplifications are justified since we
are interested in particle production at mid-rapidity. Pressure is
related to energy density with an equation of state (EoS). Our choice here is the recent
lattice EoS from Ref.~\cite{Huovinen:2009yb}.

Thermal transverse momentum spectra of hadrons are calculated with the Cooper-Frye
\cite{Cooper} method from a constant-temperature freeze-out hypersurface. We
choose the freeze-out temperature to be $T_{\text dec} = 160 (165)$~MeV for the WN (BC)
profile, which leads to a reasonable agreement with the measured pion spectra in 
$\sqrt{s_{NN}}=200$~GeV Au+Au collisions at RHIC. In each event
we have sampled the hadrons from the obtained thermal $p_T$ spectra, and used PYTHIA
\cite{Sjostrand:2006za} to do the resonance decays as explained in
Ref.~\cite{Holopainen:2010gz}.

When we study the angular dependence of the hard partons, we must define the
reference plane for each event. Theoretically the simplest choice would
be the reaction plane defined by the impact parameter and beam axis. However,
in the experiments the impact parameter is not known. Thus we follow
Ref.~\cite{PHENIX_RAA_Phi} and use the event plane as our reference plane\footnote{Note that the reference plane called "reaction plane" in Ref.~\cite{PHENIX_RAA_Phi} corresponds to our event plane.}.

The event flow vector ${\bf Q}_2$ for each event is defined as
\begin{equation}
  {\bf Q}_2 \equiv \sum_i ( P_{Ti} \cos(2\phi_i),  P_{Ti} \sin(2\phi_i) ),
\end{equation}
where we sum over all particles in the event. The event plane is then
calculated from the azimuthal angle of the vector ${\bf Q}_2$ as 
\begin{equation}
  \psi_{\rm EP} = \frac{ \arctan ( Q_y / Q_x ) }{2},
\end{equation} 
where arctan is calculated in the correct quadrant. Since in our approach
we have only a finite number of particles in the final state, the event plane
fluctuates around the ''true'' event plane, which would be determined from
infinitely many particles. To eliminate these fluctuations we make 200 events
from each hydro run and determine the final event plane as the average of
these 200 event planes. This final event plane is now very close to the ''true''
event plane and we can safely neglect the effects from the event plane
fluctuations.

The results for final hadron $P_T$ spectra and elliptic flow from this EbyE hydrodynamical
model with the WN profile can be found in Ref.~\cite{Holopainen:2010gz}. The EoS
used in Ref.~\cite{Holopainen:2010gz} is slightly different, but as shown in
Ref.~\cite{Huovinen:2009yb}, the differences between these EoS with regard  
to the final state hadronic observables are very small.

For comparison purposes, we also define a "smooth initial condition" by averaging over 10 000 MCG configurations as described above. We do not perform any rotation of the configurations used in the averaging process, thus the event plane of the smooth initial condition scenario equals to the reaction plane. 

\section{Energy loss modelling in a lumpy background}
\label{sec3}

When we change from a smooth medium to a medium evolving from fluctuating initial conditions, we qualitatively expect to
be sensitive to the following four main effects on the energy loss of leading partons:

\begin{itemize}

\item
In a medium with fluctuations in the initial state, the density distribution is not smooth but divides into clusters with  densities lower and higher than the average. Were the suppression of high $P_T$ partons a phenomenon linear in density, this would not matter, but in the observed region $R_{AA}$ is a non-linear function of the medium density where the effect of reducing the density is stronger than the effect of increasing the density by the same factor\footnote{See e.g. \cite{SysHyd1} for an explicit computation --- a simple argument states that $R_{AA}$ is a function bounded from below by zero and that complete suppression of partons corresponds to the limit of infinite medium density, i.e. a strong increase of density is needed to push the suppression below 0.2}. From these considerations, we can expect the suppression to weaken when fluctuations (corresponding to a re-distribution of the smooth medium density) are taken into account, i.e. we expect $R_{AA}$ to increase.

\item
Regions of high and low density in the initial state lead early in the hydrodynamical evolution to sharp pressure gradients, which immediately start to smooth out the fluctuations. This process implies an irregular flow field in which the local direction of the velocity field can be quite different from the late time radial plus elliptic flow pattern. In particular, the flow vector may initially point inward to the medium center. To the degree that flow influences the strength of energy loss, this effect should be visible in the models. {\em A priori}, the sign of this effect is unknown and needs to be determined in a calculation.

\item
In a fluctuating initial state, clusters of high density are distributed event by event around the vertices of binary collisions (this remains true even if wounded nucleon scaling is assumed, as the position of a wounded nucleon also implies a binary collision at this transverse position). However, since hard parton production takes place in binary collisions, there is a marked correlation between the initial medium density distribution and the hard parton production point: Hard partons tend to be produced in regions where the matter density is higher than average. This increases the suppression and is expected to lead to a decrease of $R_{AA}$ when fluctuations are taken into account.

\item
In principle, as discussed in \cite{Holopainen:2010gz}, the event plane angle differs from the reaction plane angle by a certain amount in each event. If the event averaging is done relative to the reaction plane rather than the event plane, a reduction of the 
angular modulation of $R_{AA}$ may result. Therefore, in the averaging procedure, we make use of the event plane information.

\end{itemize}

Note that the hydrodynamical evolution itself smoothes the fluctuations over time, i.e. all effects listed above become weaker in the later stages of the medium evolution. On the other hand, energy loss has a characteristic length/time dependence inherent to the underlying physics model. Thus it is not expected that a radiative energy loss model and an elastic energy loss model probe the initial state fluctuations in the same way. Therefore, we investigate the effect of fluctuations in the medium density in two different models for leading-parton energy loss. 

In order to illustrate the effect of fluctuations most clearly, we use the following strategy: We compute the dependence of $R_{AA}(P_T = 10 \, \text{GeV},\phi)$ on the angle $\phi$ of the outgoing high-$P_T$ hadrons with the event plane in 0-10\% central collisions for a number $N$ of different events with fluctuating initial conditions. For each single event, we average over the possibility that the hard parton may emerge from any of the binary collision points associated with the event. This could be called "ideal tomography", as it contains the maximum possible tomographical information that could be \emph{in principle} obtained from the event (in practice, however, it is astronomically unlikely to find a large number of high $P_T$ partons in any single event). This results in an $R_{AA}^i(\phi)$ for the event $i$ which illustrates the intra-event fluctuations as different directions lead to paths traversing or passing by dense regions in the medium. We compare different $R_{AA}^i(\phi)$ to illustrate the magnitude of the inter-event fluctuations, and define the final result as the average over the $N$ distinct events as $R_{AA}(\phi) = \sum_{i=0}^N R_{AA}^i(\phi)/N$. We determine the medium parameters characteristic for the energy loss model by the requirement that $\int_0^{2\pi} R_{AA}(\phi)/(2\pi) = R_{AA}^{\rm smooth}$. In other words, we require that the result reproduces the angular average of a calculation with a smooth initial condition
in which the medium parameters have been adjusted to describe the data at the same $P_T$. We find no significant $P_T$ dependence of the fluctuation effect, which is expected as the $P_T$ dependence of $R_{AA}$ is not very sensitive to the energy loss model \cite{gamma-h}. 

We consider three different scenarios for the fluctuations.  In the standard scenario, the initial bulk matter distribution follows the WN scaling and the characteristic size parameter $\sigma$ of the fluctuations is 0.4 fm. In order to test the hypothesis that we recover the smooth initial condition result for large-sized fluctuations, we study a second scenario where $\sigma=0.8$~fm. Finally, in order to test the sensitivity to the assumption of the WN geometry, we 
repeat the study by using a BC-distributed energy density with $\sigma=0.4$~fm.

\subsection{Radiative energy loss}

We use the Armesto-Salgado-Wiedemann (ASW) formalism in the formulation of \cite{QuenchingWeights} to compute radiative energy loss. From a given binary collision vertex and with a given orientation with respect to the event plane, we compute the two line-integrals along the medium trajectory

\begin{equation}
\label{E-lineintegral1}
  Q_s({\bf r}_0,\phi) \equiv \langle \hat{q} L \rangle 
                          = \int d\xi\, \hat{q}(\xi) 
\end{equation}

and

\begin{equation}
\label{E-lineintegral2}
  \omega_c ({\bf r}_0,\phi)= \int d\xi\, \xi\, \hat{q}(\xi).
\end{equation}

We assume that the transport coefficient $\hat{q}$ is related to the local medium energy density $\epsilon$ and the hydrodynamically computed local flow velocity $\rho$ \cite{SysHyd1}
\begin{equation}
\label{E-qhat}
  \hat{q}(\xi) = K_{\rm med} \cdot 2 \cdot \epsilon^{3/4}(\xi) 
                 \bigl(\cosh \rho(\xi) - \sinh \rho(\xi) \cos\alpha(\xi)\bigr),
\end{equation}
where $\alpha$ is the angle between the parton trajectory and flow vector, and $K_{\rm med}$ is an adjustable parameter which is determined by requiring that the model describes the angular-integrated $R_{AA}$ in central collisions.

Using the numerical results of \cite{QuenchingWeights}, $Q_s$ and $\omega_c$ determine the energy loss probability density $P(\Delta E)$ which we average at a given $\phi$ over all binary collision vertices in a given event. As explained in detail in \cite{Scaling}, this procedure requires that $\hat{q}(\xi)$ along the parton trajectory can be written in the form $~1/\tau^{\alpha}$. While this is obviously realized in the Bjorken model of only longitudinal scaling flow,  it is not self-evident that this condition is met in a hydrodynamical expansion with initial state fluctuations and buildup of transverse flow. We have however checked for a large number of trial trajectories that the leading behaviour is given by $~1/\tau^{\alpha}$ (with $\alpha$ dependent on the trajectory and event) and that the deviations are of order of few percent. This can be understood from the fact that the hydrodynamical evolution smoothes out strong density fluctuations very early already. 

We denote the resulting average probability density as $\langle P(\Delta E)\rangle_\phi$. We calculate the momentum spectrum of hard partons in leading order (LO) perturbative QCD  (explicit expressions are given in 
\cite{SysHyd1} and references therein). The medium-modified perturbative production of hadrons at an angle $\phi$ can then be computed from the expression

\begin{equation}
\label{E-LOpQCD}
  \frac{d\sigma_{\rm med}^{AA\rightarrow h+X}}{d\phi} \negthickspace 
  \negthickspace 
  = \sum_f \frac{d\sigma_{\rm vac}^{AA \rightarrow f +X}}{d\phi} 
    \otimes \langle P(\Delta E)\rangle_\phi 
    \otimes D_{f \rightarrow h}^{\rm vac}(z, \mu_F^2),
\end{equation} 
where $D_{f \rightarrow h}^{\rm vac}(z, \mu_F^2)$ is the fragmentation function 
with a momentum fraction $z$ at a scale $\mu_F^2$ set the hadronic $P_T$ \cite{KKP}. From this we 
compute the nuclear modification function $R_{AA}^i$  for the event $i$  as a function of the particle's angle with respect to the event plane as 

\begin{equation}
  R_{AA}^i(P_T,y,\phi) = \frac{dN^i_{AA}/dP_Tdy d\phi}
                            {\langle N_{\rm BC}/\sigma_{NN}\rangle\,d\sigma^{pp}/dP_Tdy d\phi},
\end{equation}
where $\langle N_{\rm BC} \rangle$ is the average number of binary collisions at a given centrality.

\subsection{Elastic energy loss}

We model the elastic energy loss of a hard parton as discussed in Ref.~\cite{Auvinen:2009qm}, by incoherent partonic $2\rightarrow$ 2 processes in pQCD, with scattering partners sampled from the medium. Our simulation of energy losses of high-energy partons in the produced QCD matter is based on the scattering rate for a high-energy parton of a type $i$, 
\begin{equation}
\label{totgamma}
\Gamma_i (p_1,u(x),T(x)) = \sum_{j(kl)} \, \Gamma_{ij\rightarrow kl}(p_1,u(x),T(x)),
\end{equation}
where we account for all possible partonic processes $ij\rightarrow kl$ by summing over all types of collision partners, $j=u,d,s,\bar u, \bar d, \bar s,g$ in the initial state, and over all possible parton type pairs $(kl)$ in the final state. In general, the scattering rate depends on the frame, and in particular on the high-energy parton's 4-momentum $p_1$, on the flow 4-velocity $u(x)$ and on the temperature $T(x)$ of the fluid at each space-time location $x$. 

In the local rest-frame of the fluid, we can express the scattering rate as follows \cite{Auvinen:2009qm}:

\begin{equation}
\label{scattrate}
\Gamma_{ij\rightarrow kl} = \frac{1}{16\pi^2E_1^2}\int_{\frac{m^2}{2E_1}}^{\infty}dE_2f_j(E_2,T) \int_{2m^2}^{4E_1E_2}ds [s\sigma_{ij\rightarrow kl}(s)].
\end{equation}

Here $E_1$ is the energy of the high-energy parton $i$ in this frame and $E_2$ is the energy of the thermal particle $j$ with a distribution function $f_j(E_2,T)$, which is the Bose-Einstein distribution for gluons and the Fermi-Dirac distribution for quarks. The scattering cross section $\sigma_{ij\rightarrow kl}(s)$ depends on the standard Mandelstam variable $s$. A thermal-mass-like overall cut-off scale $m=g_sT$ is introduced in order to regularize the singularities appearing in the cross section when the momentum exchange between partons approaches zero. Here $g_s$ is the strong coupling constant, which we keep fixed with momentum scale in this work.

To initiate the hard massless parton of a type $i$ in each event, we sample the partonic $p_T$ from the LO pQCD single-parton production spectrum  $d\sigma/dp_Tdy_i$ \cite{Eskola:2002kv} at $y_i=0$ in the range  $p_{T{\rm min}}\leq p_T\leq \sqrt s/2$. For the parton distribution functions (PDFs), we use the CTEQ6L1 set \cite{CTEQ}. The nuclear effects to the PDFs \cite{NPDF,EKS98,EPS09} are considered small in comparison with the final state medium interactions and thus neglected. The initial rapidity \(y_i\) is randomly generated in the range $|y_i|\le y_{\rm max}$ from a flat distribution. This fixes the hard-parton energy \(E\) and polar angle \(\theta\) of its momentum vector ${\bf p}=(p_x,p_y,p_z)$. The azimuth angle \(\phi\), defined with respect to the event plane, is evenly distributed between \([0,2 \pi]\). 

The hard parton is assumed to start interacting with the medium at the initial longitudinal proper time \(\tau_0\) of our hydrodynamical model. Since in the c.m. frame of the colliding nuclei all hard partons are produced in the Lorentz-contracted overlap region at $z\approx 0$, the longitudinal position at later times (before the first collision at $\tau\ge \tau_0$) is assumed to be determined by the longitudinal momentum only. The initial time and longitudinal coordinates for the hard parton are thus \(t_0 = \tau_0 \cosh{y_i}\) and \(z_0 = \tau_0 \sinh{y_i} \). The coordinates on the transverse plane in the beginning of the simulation are then $x_0=x_i+\frac{p_x}{E}t_0$ and $y_0=y_i+\frac{p_y}{E}t_0$, where the parton position $(x_i,y_i)$ at $t=0$ is sampled from the nuclear overlap function for the smooth case. With the fluctuating initial conditions, the parton starting point is randomly sampled from the set of binary collision vertices in each event (see e.g. Fig.~\ref{F-RAA-hydro}).

The hard parton propagates through the plasma in small time steps \(\Delta t\), during which we propagate the parton in position space. The probability for {\em not} colliding in this time interval is assumed to be given by the Poisson distribution $e^{-\Gamma_i \Delta t}$, where \(\Gamma_i\) is the total scattering rate \eqref{totgamma} for the hard parton of the type $i$. For small enough \(\Delta t\), we can assume that there will be at most one collision. As we calculate the scattering rates \eqref{totgamma} in the local rest frame of the quark-gluon plasma fluid element, the time step \(\Delta t\) is also boosted to the same frame. Should a scattering happen, the probability $P_{ij\rightarrow kl}$ for a given type of scattering process is determined by the ratios of the partial scattering rates \eqref{scattrate} to the total scattering rate \eqref{totgamma}. After scattering, the final state parton with highest energy is chosen as the new hard parton to be propagated further, for which we repeat the procedure outlined above with the next timestep. 

We take into account the system's gradual transformation from quark-gluon plasma to hadron gas by using an effective temperature $T_{eff}=\left(\frac{30}{g_Q \pi^2}\epsilon\right)^{1/4}$, where \(g_Q=g_g + \frac{7}{8}2N_fg_q = \frac{95}{2}\) is the quark-gluon plasma degrees of freedom with gluon and quark DOF being $g_g=16$ and $g_q=6$, respectively, and number of quark flavors $N_f=3$. We always assume there is no significant interaction between the high-energy parton and the fully hadronic medium, and thus no collisions happen in regions with temperature below $T_{\rm dec}$.

The outcome of the procedure described above is a medium-modified distribution of high-energy partons, $\frac{dN^{AA \rightarrow f+X}}{dp_T dy}$. Analogously with the radiative energy loss case \eqref{E-LOpQCD}, the obtained partonic distribution is convoluted with the fragmentation function $D_{f \rightarrow h}^{\rm vac}(z, P_T^2)$ in order to calculate the nuclear modification factor $R_{AA}^i(P_T,y,\phi)$. 

We average the obtained $R_{AA}^i(P_T,y,\phi)$ across the rapidity window $[-0.35,0.35]$ which corresponds to the PHENIX acceptance. Due to the non-eikonal propagation of the hard partons in the simulation, the initial rapidity window is set by choosing $y_{\rm max} = 0.7$ in order to account for all the possible partons falling into the final rapidity window \cite{Auvinen:2009qm}. To achieve roughly the right amount of nuclear modification in the 0-10\% centrality bin and to emulate also the incoherent higher-order processes, the value of strong coupling constant is chosen to be $\alpha_s = 0.5$. For the initial conditions in our elastic energy loss simulations, the BC profile with the width parameter $\sigma=0.4$~fm are used exclusively.

\section{Results}
\label{sec4}

\subsection{Radiative energy loss, central collisions}

\begin{figure}[htb]
\epsfig{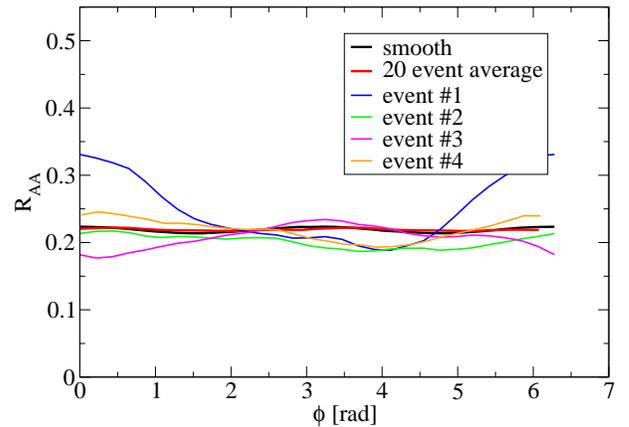}
\caption{\label{F-RAA-rad-20evt} The nuclear suppression factor $R_{AA}$ for central 200 AGeV Au-Au collisions at $P_T=10$ GeV as a function of the angle of outgoing hadrons with the event plane shown with smooth initial conditions, for four different events with fluctuating initial conditions averaged over all binary collision vertices ('ideal tomography') and averaged over 20 such events.}
\end{figure}

In Fig.~\ref{F-RAA-rad-20evt} we show one of our main results for radiative energy loss, i.e. the angular dependence of $R_{AA}$ both for a number of events with fluctuating initial state geometry in the standard scenario and averaged over 20 different events, compared with the result for smooth, averaged initial conditions.

It is readily apparent from the result that there are large inter-event fluctuations as well as somewhat smaller intra-event fluctuations. For this centrality class, both types of fluctuations are much larger than the angular modulation induced by the spatial anisotropy of the matter distribution for the smooth initial condition. A 20 event average captures however the normalization and partially also the angular modulation with respect to the event plane. For the standard fluctuation scenario, the value of $K_{\rm med}$ in Eq.~(\ref{E-qhat}) extracted from the data is to the limit of our statistical accuracy identical with the one from the smooth result, thus we do not observe any modification of the medium quenching power in this fluctuation scenario.

\begin{figure}[htb]
\epsfig{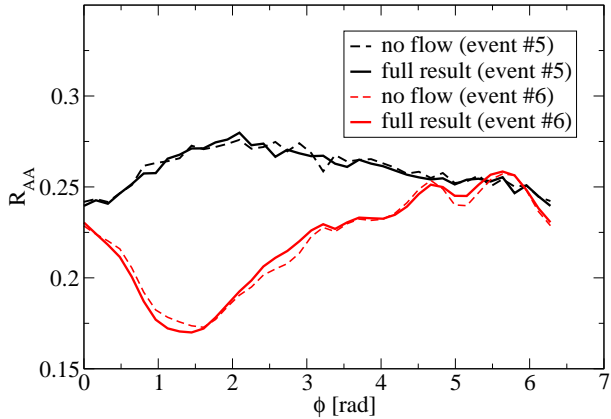}
\caption{\label{F-RAA-flow} The nuclear suppression factor $R_{AA}$ for central 200 AGeV Au-Au collisions at $P_T=10$ GeV as a function of the angle of outgoing hadrons with respect to the event plane shown for two different events with the flow velocity set to zero and evaluating the full flow boost.}
\end{figure}

We may understand this finding better by considering the individual mechanisms how fluctuations may influence the outcome in detail. In Fig.~\ref{F-RAA-flow} we show the effect of the irregular flow field by comparing the full result with a result in which the flow velocity in Eq.~(\ref{E-qhat}) has been artificially set to zero. This means however that also the effect of the mean radial flow field which is present in the smooth result has been taken out, thus we rescale $K_{\rm med}$ in Eq.~(\ref{E-qhat}) with the factor obtained from a calculation with smooth initial conditions where the flow effect is also taken out. It is readily apparent from the figure that the influence of fluctuation-induced flow on the result is small.

\begin{figure}[bht]
\epsfig{file=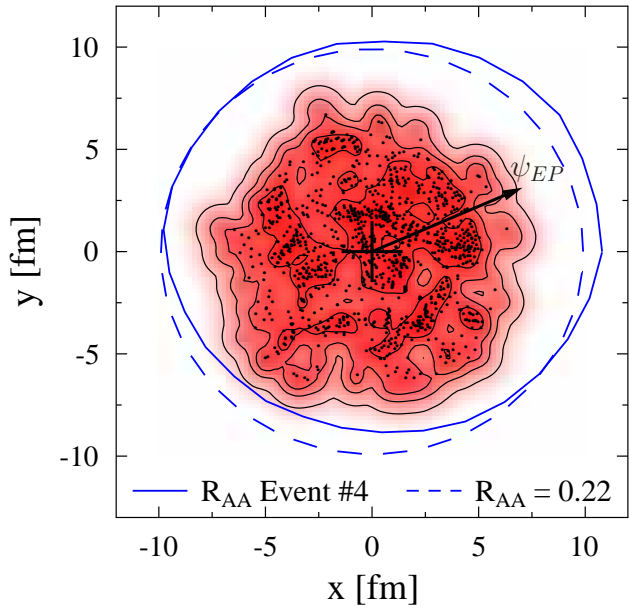, width=9cm}
\caption{\label{F-RAA-hydro} 
The initial state temperature profile, BC points and the direction of the event plane for a selected event (number 4 in Fig.~\ref{F-RAA-rad-20evt}). The constant-temperature contours are for $T=160,300,400,500$~MeV. Shown as polar plots
are also the $R_{AA}(\phi)$ for this event (solid) and a constant $R_{AA}(\phi)=0.22$ (dashed).}
\end{figure}

To demonstrate the correlation between the initial state density fluctuations and the angular behavior of $R_{AA}$,
we have plotted in Fig.~\ref{F-RAA-hydro} the initial temperature profile in
the transverse plane, the BC points and, as a  a polar plot, the calculated $R_{AA}(\phi)$ for the event number 4 of Fig.~\ref{F-RAA-rad-20evt}. We have also added a polar plot of $R_{AA}(\phi) = 0.22$ to guide the eye.
Since many of the BC points are in the hot spot in the 1st quadrant, the probability
to produce a hard parton from this region is large. Thus a significant contribution
to $R_{AA}$ comes from partons emerging from this hot spot.
If these partons are going into the direction of the event plane, they have less
medium to traverse and thus do not lose as much energy as those which are
moving into the opposite direction. Hence the $R_{AA}$ is larger near the event plane angle $\psi_{\rm EP}$.
\begin{figure}[htb]
\epsfig{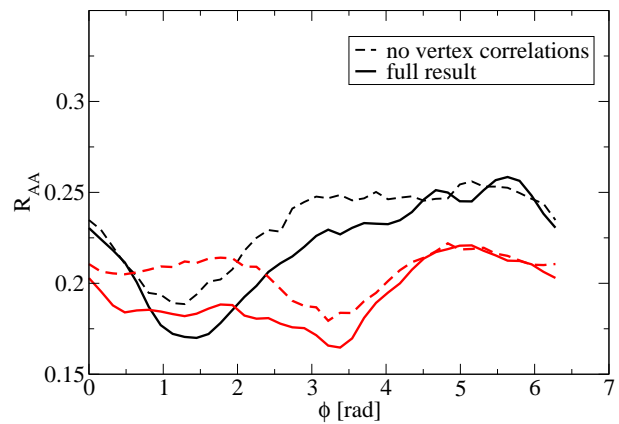}
\caption{\label{F-RAA-bin} The nuclear suppression factor $R_{AA}$ for central 200 AGeV Au-Au collisions at $P_T=10$ GeV as a function of the angle $\phi$ of outgoing hadrons with respect to the event plane shown for two different events averaged with uncorrelated random binary collision points from a smooth nuclear overlap function or binary collision vertices correlated with the event fluctuations.}
\end{figure}

In Fig.~\ref{F-RAA-bin} we show the angular dependence of $R_{AA}$ averaged over both randomly chosen initial binary collision points according to a smooth overlap distribution and averaged over the actual distribution of binary collision points of the event. It can be seen  from the figure that accounting for the correlations of binary collision points with medium density leads to increased suppression and a reduction of $R_{AA}$ of $O(10)$\% on average.

These results suggest that the decrease in suppression caused by the non-linearity of the response of $R_{AA}$ to the medium density is just compensated by the increase in suppression due to the correlation of initial vertices with dense matter regions, rendering the final result in terms of extracted medium properties almost unchanged. However, this cancellation is somewhat accidental, as will be seen in the discussion of other scenarios for the fluctuations.

\begin{figure}[htb]
\epsfig{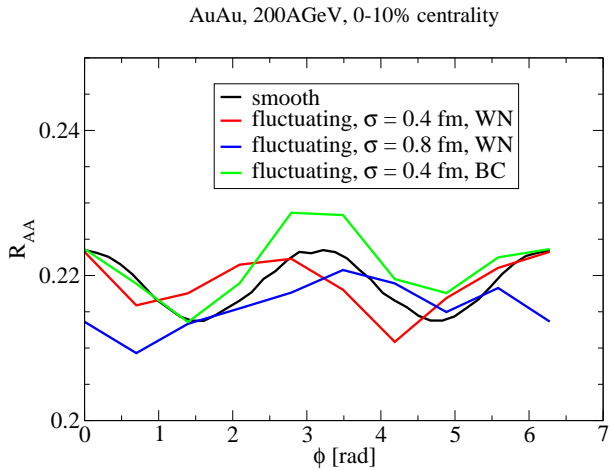}
\caption{\label{F-RAA-0_10} The nuclear suppression factor $R_{AA}$ for 0-10\% central 200 AGeV Au-Au collisions at $P_T=10$ GeV as a function of the angle $\phi$ of outgoing hadrons with respect to the event plane shown with smooth initial conditions and as 10 event average over fluctuating initial conditions in three different scenarios (see text). Note the scale of the $y$-axis.}
\end{figure}

In Fig.~\ref{F-RAA-0_10} we show the averaged angular modulation of $R_{AA}$ for all three fluctuation scenarios discussed previously, after a 10 event average where the parameter $K_{\rm med}$ in Eq.~(\ref{E-qhat}) has been adjusted to result in the best fit to the data in the 0 to 10\% centrality class, zoomed into the relevant $y$-axis region. Already a 10 event average recovers some of the angular modulation seen in a smooth event. For $\sigma = 0.4$ fm and WN geometry, the extracted $K_{\rm med}$ is, as mentioned above, the same as in the smooth case. For $\sigma = 0.8$ fm we extract a value that is, somewhat surprisingly, 18\% smaller than in the smooth case. Note that this is not a large effect, since the value of $K_{\rm med}$ may change by a factor two between different hydrodynamical evolution models \cite{SysHyd3,JetHyd}. These results suggest that fluctuations do not have a substantial effect on the extraction of medium parameters, but rather constitute an additional uncertainty.

The value of $K_{\rm med}$ in the BC geometry is 30\% larger than in the smooth case, but most of this difference can be attributed to the density profile geometry rather than the fluctuation effect. Vertices for hard parton production are distributed according to the BC geometry. If bulk matter is distributed according to the (wider) WN geometry, the mean in-medium path for hard partons is larger than if bulk matter is also distributed with the (narrower) BC geometry. Hence quenching is on average stronger for a WN geometry, requiring a larger $K_{\rm med}$ if $R_{AA}$ is required to agree with the data.

\subsection{Radiative energy loss, non-central collisions}

Let us now turn to the consequences of fluctuations for non-central collisions. Here, the angular modulation of $R_{AA}$ is much less influenced by fluctuations and already a 10 event average recovers the sinusoidal modulation with a good accuracy. This result is expected and also seen at low $P_T$ in the computation of bulk matter $v_2$.

\begin{figure}[htb]
\epsfig{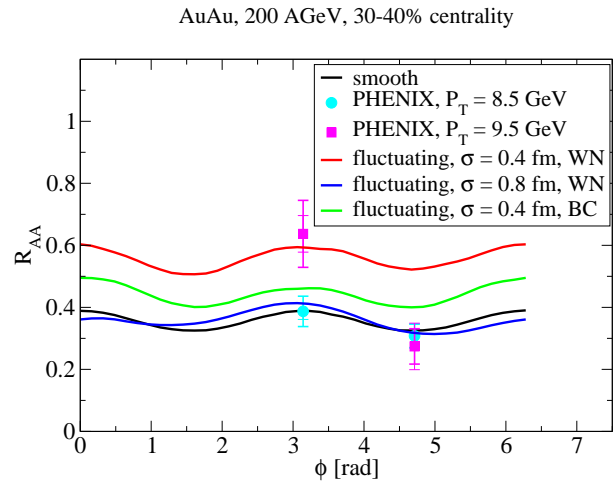}
\caption{\label{F-RAA-30_40} The nuclear suppression factor $R_{AA}$ for 30-40\% peripheral 200 AGeV Au-Au collisions at $P_T=10$ GeV as a function of the angle $\phi$ of outgoing hadrons with respect to the event plane shown with smooth initial conditions and as 10 event averages over fluctuating initial conditions in three different scenarios (see text), compared with PHENIX data \cite{PHENIX_RAA_Phi}.}
\end{figure}

Our results for the three different scenarios are shown in Fig.~\ref{F-RAA-30_40} along with PHENIX data for $P_T = 8.5$ GeV and $9.5$ GeV \cite{PHENIX_RAA_Phi}. It can be seen that the normalization of the results with fluctuating initial conditions is quite different from the smooth result, although the magnitude of the angular modulation is roughly similar. In particular, small-sized fluctuations increase the normalization above the smooth result, whereas it is reassuring that a larger spatial scale for the size of the fluctuations leads back to the smooth result.

Since the $P_T$ dependence of the data is not smooth and in particular the 9-10 GeV bin suffers from large statistical and systematic errors, it is not entirely clear if the data can rule out one scenario at this point. However, taking the $P_T$ average of the data, it certainly seems that the normalization found for small-sized fluctuations $\sigma=0.4$ fm in the WN scenario is not supported by the data. This is an interesting result, as it would allow to constrain the typical size of initial state fluctuations.

\subsection{Elastic energy loss}

\begin{figure}[htb]
\epsfig{file=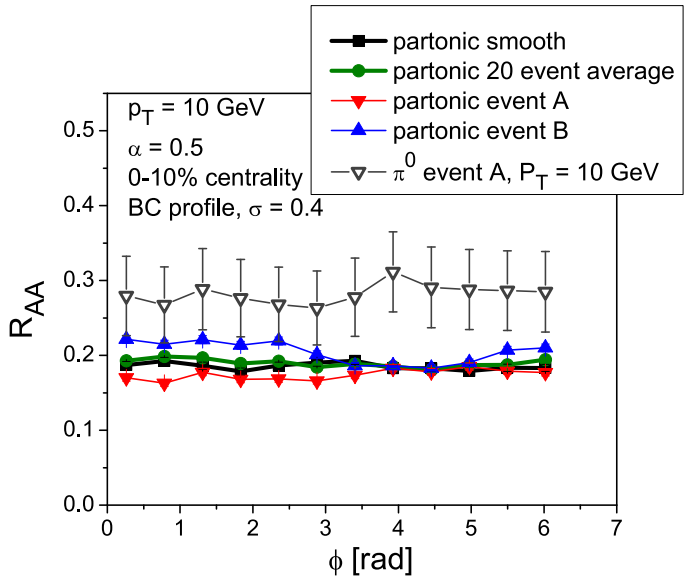, width=9cm}
\caption{\label{F-RAA-elastic-20evt} The partonic nuclear suppression factor $R_{AA}$ for central 200 AGeV Au-Au collisions at $p_T=10$ GeV as a function of the angle of outgoing partons with respect to the event plane shown for smooth initial conditions, for four different events with fluctuating initial conditions and for an average over 20 fluctuation events. Nuclear suppression factor for $\pi^0$ at $P_T=10$ GeV in one event is also displayed.}
\end{figure}

Figure \ref{F-RAA-elastic-20evt} is analogous to Fig.~\ref{F-RAA-rad-20evt}, comparing the angular dependence of $R_{AA}$ for fluctuating initial state geometry with a result for smooth initial conditions. As the individual events do not necessarily correspond to the events used in the radiative energy loss model, they are here labeled with letters instead of numbers. Also, the results are for partons instead of hadrons, as the convolution with fragmentation functions requires considerable amount of simulation data and increases the statistical uncertainties, but contributes very little to the results beyond the change in normalization, as can be seen in Fig.~\ref{F-RAA-elastic-20evt} for the event A.

\begin{figure}[htb]
\epsfig{file=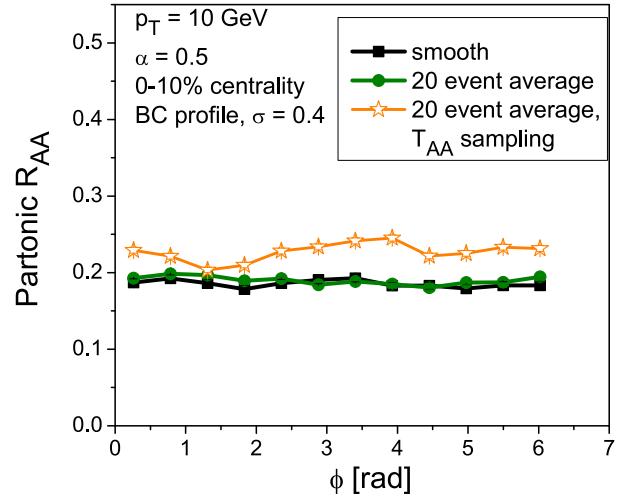, width=8cm}
\caption{\label{F-RAA-elastic-correlations} The partonic nuclear suppression factor $R_{AA}$ for central 200 AGeV Au-Au collisions at $p_T=10$ GeV as a function of the angle of outgoing partons with respect to the event plane, shown for smooth initial conditions, for an average over 20 fluctuation events where parton starting points are correlated with binary collision vertices, and for a 20-event fluctuation average with parton starting points sampled from the corresponding nuclear overlap function $T_{AA}({\bf b})$.}
\end{figure}

Notable inter-event variation was seen in the radiative energy loss scenario, and this seems to be true also with elastic energy loss. The intra-event angular variation, however, is rather weak. The average over 20 events with fluctuating initial conditions, keeping the same value $\alpha_s=0.5$ for the  fluctuating and smooth cases, equals the smooth initial condition scenario with fairly good accuracy, as was the case with the radiative energy loss model. 
This is perhaps more clearly seen in Fig.~\ref{F-RAA-elastic-correlations}, in which we also study the scenario where the starting points of high-energy partons are sampled from the nuclear overlap function $T_{AA}({\bf b})$ in a medium with fluctuating initial conditions, as opposite of using binary collision vertices. This effectively removes the local correlation between the parton starting points and the regions with high energy density in the medium. One would expect this to decrease the suppression, and indeed the nuclear modification factor increases about 20\% in this scenario.

\begin{figure}[tb]
\epsfig{file=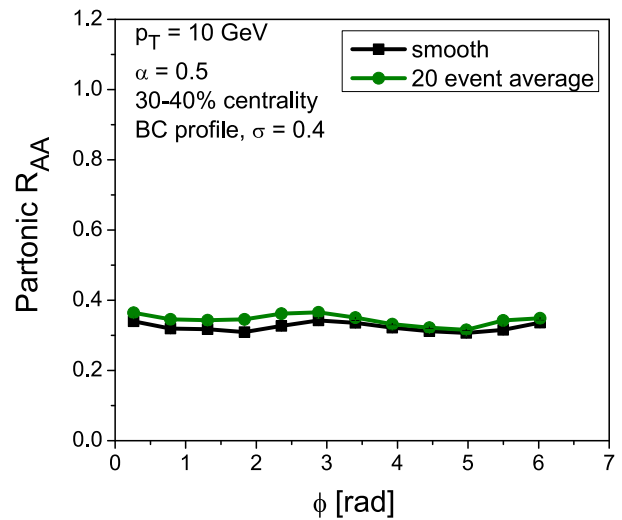, width=8cm}
\caption{\label{F-RAA-elastic-3040} The partonic nuclear suppression factor $R_{AA}$ for 30-40\% peripheral 200 AGeV Au-Au collisions at $p_T=10$ GeV of the angle of outgoing partons with respect to the event plane shown with smooth initial conditions, and averaged over 20 events with fluctuating initial conditions.}
\end{figure}

Moving to the non-central collisions, we compare Fig.~\ref{F-RAA-elastic-3040} with Fig.~\ref{F-RAA-30_40}. As in the radiative energy loss scenario, the average of the events with fluctuating initial conditions is found to be systematically above the smooth initial conditions curve. The difference between the two curves is quite small in this scenario, but enough to argue that both the radiative and the elastic energy loss models see a difference between the smooth and the fluctuating initial conditions in the peripheral case.

\section{Discussion}

We have presented a systematic study of the effects of QCD-matter density fluctuations on the angular dependence of the nuclear modfication factor $R_{AA}$, comparing a radiative and an elastic energy loss model. In general, the overall effects of fluctuations on $R_{AA}$ observables seem to be rather small. 

In central collisions, an extraction of medium parameters in the radiative energy-loss model resulted in a difference of only $~\sim 20$ \% for the different fluctuation size-scales studied. For the default fluctuation scenario, the 20-event average reproduced both the normalization and the angular modulation of $R_{AA}$ of the smooth scenario with unchanged medium parameter $K_{\rm med}$. This is due a cancellation of the increased suppression due to the correlation of parton production points with the hot spots, and a decrease of the suppression due the nonlinearity of $R_{AA}$ with respect to the medium density. For non-central collisions, however, we find that this cancellation is incomplete, and as a result the angular-integrated $R_{AA}$ depends visibly on the fluctuation size, whereas the amplitude of the angular modulation is roughly the same as in the smooth scenario. This observation may help in constraining the size-scale of the initial QCD-matter density fluctuations.

The results from the radiative and the elastic energy loss models seem to agree on qualitative level, even though the weak sensitivity of the elastic energy loss model to the angle-dependent observables already seen in \cite{Auvinen:2010yt} is clearly seen also in this research. In the central collisions, no difference is seen between the fluctuating and the smooth initial conditions when the average over 20 events has been taken. However, as in the case of radiative energy loss, in the peripheral collisions the fluctuating conditions appear to produce somewhat smaller suppression compared to the smooth background.

Interestingly, while Fig.~\ref{F-RAA-30_40} would suggest a fairly large size for the fluctuations (the $\sigma=0.8$~fm case agrees better with the PHENIX data), the recent study on the effects of fluctuations on thermal photon production \cite{Chatterjee:2011dw} would favor a smaller fluctuation size. Thus a combined analysis has a strong potential to constrain the initial state fluctuation scale size and thus to identify the underlying physics mechanisms for the bulk QCD-matter production.

\begin{acknowledgments}
T.R. is supported by the Academy researcher program of the Academy of Finland (Project No. 130472) and Academy Project 133005. H.H. gratefully acknowledges the financial support from the national Graduate School of Particle and Nuclear Physics and J.A. the grant from the Jenny and Antti Wihuri Foundation. We acknowledge CSC -- IT Center for Science in Espoo, Finland, for the allocation of computational resources.
 
\end{acknowledgments}

\end{document}